# Wick's Theorem And a New Perturbation Theory Around The Atomic Limit of Strongly Correlated Electron Systems

JAN BRINCKMANN

Institut f. Festkörperphysik, Technische Hochschule Darmstadt
Hochschulstr. 6, D-64289 Darmstadt, Germany

### Abstract

A new type of perturbation expansion in the mixing $V$ of localized orbitals with a conduction-electron band in the $U \to \infty$ Anderson model is presented. It is built on Feynman diagrams obeying standard rules. The local correlations of the unperturbed system (the atomic limit) are included exactly, no auxiliary particles are introduced. As a test, an infinite-order ladder-type resummation is analytically treated in the Kondo regime, recovering the correct energy scale. An extension to the Anderson-lattice model is obtained via an effective-site approximation through a cumulant expansion in $V$ on the lattice. Relation to treatments in infinite spatial dimensions are indicated.

**Correspondence to:**   J. Brinckmann
　　　　　　　　　　　　Inst. für Festkörperphysik,
　　　　　　　　　　　　Technische Hochschule Darmstadt
　　　　　　　　　　　　Hochschulstr. 6,
　　　　　　　　　　　　D-64289 Darmstadt
　　　　　　　　　　　　Fed. Rep. of Germany
　　　　　　　　　　　　**Fax.:** ++49-(0)6151 / 163681
　　　　　　　　　　　　**e-mail:** jan@spy.fkp.physik.th-darmstadt.de

The large Coulomb repulsion $U$ on localized $f$- or $d$-orbitals in strongly correlated electron systems is the source of exciting many-body effects and at the same time responsible for great technical difficulties: in a theoretical description based on Anderson- or Hubbard-type models, the most suitable starting point uses local (atomic) many-body states, which are perturbed by an intersite hopping $t$ or a hybridisation $V$ with delocalized band states. The atomic limit is apparently *not* given by a quadratic (i.e. one-particle) Hamiltonian $H_0$, and the use of a conventional Feynman-diagram technique in a perturbation expansion with respect to $t$ or $V$ poses an yet unsolved problem.

In 'slave-boson' techniques Wick's theorem is recovered via auxiliary particles, but for the price of a constraint [1, 2] to be maintained. It leads, if taken strictly [3], to the unconventional time-ordered perturbation theory [4, 5]. Other approaches [6, 7] start from commutation relations of Hubbard operators [8, 5] and aim at an adequate 'Wick's theorem', which appears as a recurrance relation and does not provide with strict diagram rules [7].

We will show in the following for the $U \to \infty$ Anderson-impurity model [9] the possibility of an expansion in the hybridisation $V$ using conventional Feynman diagrams. The strong local correlations are kept exactly with no need of 'slave' particles. As a test case for the new approach the Kondo problem is reconsidered, and finally an extension to the Anderson-lattice model is discussed.

With Hubbard transfer operators [8] $X^{\sigma\sigma} = (1 - n^f_{-\sigma})n^f_\sigma$, $X^{0\sigma} = (1 - n^f_{-\sigma})f_\sigma$, $X^{\sigma 0} = (X^{0\sigma})^\dagger$, $n^f_\sigma = f^\dagger_\sigma f_\sigma$ the $U \to \infty$ model is formulated as

$$H = \sum_{\boldsymbol{k},\sigma} \varepsilon_{\boldsymbol{k}\sigma} c^\dagger_{\boldsymbol{k}\sigma} c_{\boldsymbol{k}\sigma} + \sum_\sigma \varepsilon^f_\sigma X^{\sigma\sigma} + \sum_{\boldsymbol{k},\sigma} \left( V_{\boldsymbol{k}} X^{\sigma 0} c_{\boldsymbol{k}\sigma} + V^*_{\boldsymbol{k}} c^\dagger_{\boldsymbol{k}\sigma} X^{0\sigma} \right) \ . \quad (1)$$

The effect of $X$-operators on Fock-space states $|\alpha; N_c\rangle$, characterized by an atomic $f$-configuration $\alpha \in \{0, \uparrow, \downarrow, 2\}$ and some set $N_c$ of conduction-band-occupation numbers, differs from that of canonical $f$-operators $n_\sigma^f$, $f_\sigma$, $f_\sigma^\dagger$ through the projecting prefactors: these allow for non-vanishing transitions only in the subspace without local double occupancy, $X^{\sigma\sigma}|\sigma';N_c\rangle = |\sigma;N_c\rangle\delta_{\sigma',\sigma}$, $X^{0\sigma}|\sigma';N_c\rangle = |0;N_c\rangle\delta_{\sigma',\sigma}$, $X^{\sigma 0}|0;N_c\rangle = |\sigma;N_c\rangle$, $\sigma = \uparrow,\downarrow$. The trace in e.g. the partition function $Z = \text{Tr}'[e^{-\beta H}]$ includes only this restricted Hilbert space (denoted by the prime) i.e. matrixelements like $\langle M;N_c|e^{-\beta H}|M;N_c\rangle$ with $M = 0,\sigma$. The $X$-operators create intermediate states according to the allowed transitions above. Since the trace offers only admissible local configurations the prefactors in $X^{\sigma\sigma}$ and $X^{0\sigma}$ are actually inoperative as long as the creation of a double occupancy is still prevented in keeping $X^{\sigma 0}$ as it is: the replacement of $X^{\sigma\sigma}$ by $n_\sigma^f$ and $X^{0\sigma}$ by $f_\sigma$ does not alter the transitions occuring in the matrixelement. The term under the primed trace may therefore be simplified, leading to $Z = \text{Tr}'[e^{-\beta\widetilde{H}}]$ with

$$\widetilde{H} = \widetilde{H}^0 + \sum_{\bm{k},\sigma}\left(V_{\bm{k}}X^{\sigma 0}c_{\bm{k}\sigma} + V_{\bm{k}}^*c_{\bm{k}\sigma}^\dagger f_\sigma\right) \quad , \quad \widetilde{H}^0 = \sum_{\bm{k},\sigma}\varepsilon_{\bm{k}\sigma}c_{\bm{k}\sigma}^\dagger c_{\bm{k}\sigma} + \sum_\sigma \varepsilon_\sigma^f f_\sigma^\dagger f_\sigma \; .$$

In general this argument [10] holds for any matrixlement

$$\langle M';N_c'|\mathcal{O}(\{X^{\sigma 0},X^{0\sigma},X^{\sigma\sigma}\})|M;N_c\rangle = \langle M';N_c'|\mathcal{O}(\{X^{\sigma 0},f_\sigma,f_\sigma^\dagger f_\sigma\})|M;N_c\rangle \tag{2}$$

of any operator $\mathcal{O}$ in the subspace $M,M' = 0,\uparrow,\downarrow$ involved in the $U \to \infty$-problem. Note, the matrix of a Hermiteian operator like $\mathcal{O} = e^{-\beta H}$ remains in this subspace Hermiteian.

The local $f$-propagator $F_{\sigma,\sigma'}(\tau,\tau') = -\langle\mathcal{T}\{X^{0\sigma}(\tau)X^{\sigma' 0}(\tau')\}\rangle$ can now be written as $F_{\sigma,\sigma'}(\tau,\tau') = -\text{Tr}'[e^{-\beta\widetilde{H}}\mathcal{T}\{f_\sigma(\tau)X^{\sigma' 0}(\tau')\}]/Z$ with (imaginary time) Heisenberg operators involving $\widetilde{H}$ according to (2). It still requires a reduced trace (Tr'), the quadratic unperturbed Hamiltonian $\widetilde{H}^0$ alone is not sufficient in order to get a Feynman-type expansion instead of (time ordered) Goldstone diagrams [11]. But since $X^{\sigma' 0}(\tau') = (f_{-\sigma'}f_{-\sigma'}^\dagger f_{\sigma'}^\dagger)(\tau')$ eliminates all matrixelements $\langle 2;N_c|\ldots|2;N_c\rangle$, we may enlarge the trace to the full Fock space and are left with a conventional interacting two-particle propagator:

$$F_{\sigma,\sigma'}(\tau,\tau') = -\frac{\widetilde{Z}}{Z} <\mathcal{T}\{f_\sigma(\tau)f_{-\sigma'}(\tau'+0_+)f^\dagger_{-\sigma'}(\tau')f^\dagger_{\sigma'}(\tau')\}>^{\widetilde{H}} . \quad (3)$$

Wick's theorem is now applicable and an expansion in connected diagrams is formulated as usual (see, e.g. [12]): a diagram of order $V^{2n}$ consists of $n$ simple hybridization vertices (Fig. 1 top) and $n$ two-particle vertices (Fig. 1 bottom). These are linked by bare conduction-electron lines $G^0_{\mathbf{k},\sigma}(i\omega_l) = 1/[i\omega_l - \varepsilon_{\mathbf{k}\sigma}]$ (full lines) and bare 'unprojected' $f$-propagators $\widetilde{G}^0_\sigma(\tau,\tau') = - <\mathcal{T}\{f_\sigma(\tau)f^\dagger_\sigma(\tau')\}>^{\widetilde{H}_0} \leftrightarrow \widetilde{G}^0_\sigma(i\omega_l) = 1/[i\omega_l - \varepsilon^f_\sigma]$ (dashed lines). Its sign $(-1)^{n_c}$ is given by the number of fermion loops $n_c$, and each internal frequency $i\omega_n$ is accompanied by a factor $e^{-i\omega_n 0_+}$ (the exponent differs in sign from standard rules!).

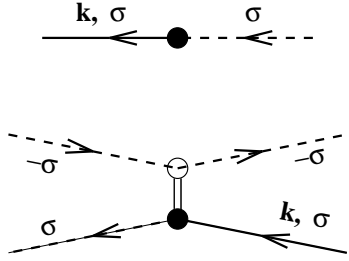

Figure 1: *Vertices resulting from the application of Wick's theorem. Spin, frequency and direction of arrows are conserved. Dots represent matrix elements $V_k$, $V_k^*$. The two-particle vertex (bottom) permits an f-creation only if a hole of opposite spin is present.*

The skeleton representation of the physical $f$-Green's function (3) shown in Fig. 2 involves fully renormalized 'unprojected' Green's functions $\widetilde{G}_\sigma(i\omega_l)$ (dashed double lines) and an irreducible vertex function $\Gamma^2$ [12]. A prefactor appears through

$$<X^{00} + X^{\sigma\sigma}> \equiv <1 - n^f_{-\sigma}> = \frac{1}{Z}\mathrm{Tr}'[e^{-\beta\widetilde{H}}(1-n^f_{-\sigma})] = \frac{\widetilde{Z}}{Z}<1 - n^f_{-\sigma}>^{\widetilde{H}} ,$$

it already reveals the exact spectral weight according to the sum rule $\int_{-\infty}^{\infty} d\omega\, \rho_\sigma^f(\omega) = <X^{00} + X^{\sigma\sigma}>$, $\rho_\sigma^f(\omega) = -\frac{1}{\pi}\mathrm{Im}F_{\sigma,\sigma}(\omega + i0_+)$. The two fermion loops emerging from contracting three of the four external time labels of (3) contain the overall sign. In the one-particle part (Fig. 2, first term) the loop also cancels the prefactor's denominator $<1 - n_{-\sigma}^f>^{\widetilde{H}}$.

$$F_{\sigma,\sigma}(i\omega_l) = \frac{<X^{00} + X^{\sigma\sigma}>}{<1 - n_{-\sigma}^f>^{\widetilde{H}}} \cdot \left\{ \text{[diagram 1]} + \text{[diagram with } \Gamma^2\text{]} \right\}$$

Figure 2: *Skeleton representation of the physical f-Green's function (see text). Conservation of incoming frequency and spin is maintained by a dummy vertex.*

The Kondo problem is now considered within the perturbation scheme formulated above in order to identify important diagram types. The simplest reasonable result for the $f$-spectral function $\rho^f(\omega)$ at low energies $\omega$ is obtained by infinite-order resummation of spin-flip-scattering processes [13], thus it may be expected that a simple mean-field (MF) theory is insufficient: based on a neglect of $\Gamma^2$ (Fig. 2, second term) and the self-consistent treatment of $\widetilde{G}(i\omega_l)$ in lowest order (spin-degenerate) skeleton-self energy [14] only a slight shift of the bare $f$-level $\varepsilon^f$ shows up. In going further, the analytical solution [15] of Bethe-Salpeter equations for the particle–hole (ph) channel and for the particle–particle (pp) channel, which contains the important spin-flip-ladder diagrams, produces via $\Gamma^2$ the Abrikosov-Suhl resonance at low

temperature $T$

$$\rho^f(\omega)\Big|_{\substack{T\to 0 \\ |\omega|\ll D}} = <X^{00} + X^{\sigma\sigma}> \frac{\pi T_A}{\Delta}\delta(\omega - T_A) \ , \ T_A = D\exp\left(\frac{\pi\varepsilon^f}{2\Delta}\right) \ ,$$

including the proper energy scale $T_A$ [13, 16]. The constant background due to the (ph)-channel has been left out here. $D$ is half the width of a flat conduction band, $\Delta \equiv \pi|V|^2/2D$ at $V_{\boldsymbol{k}} = V$.

Our result compares very well with analytical non-self-consistent resummation in time-ordered perturbation theory [17, 18]: in setting $<X^{00} + X^{\sigma\sigma}>\approx \frac{1}{N} = \frac{1}{2}$ the spectrum of [19] is exactly obtained.

An extension of the method to the $U \to \infty$ Anderson-lattice model, however, remains difficult. The operator replacement (2) still holds for each lattice site $\boldsymbol{R}_\mu \equiv \mu$ (as $X^{0\sigma}_\mu \to f_{\mu\sigma}$ etc.), but in the local $f$-Green's function $F_{\mu\mu,\sigma}(i\omega_l)$ double occupancies on all sites other than $\mu$ do contribute to an enlarged trace. If the replacement and trace-enlargement steps are performed after expanding $F_{\mu\mu,\sigma}(i\omega_l)$ in $V$ first, a bookkeeping of lattice sites involved turns out to be necessary for each diagram, thus a linked-cluster expansion required to get an intensive Green's function is prevented. The well known excluded-volume problem [5] showing up here can be treated by way of self-consistent resummations of lattice processes to an effective Anderson-impurity model [20, 21, 22], a concept closely related to recent treatments of lattice models in infinite spatial dimensions [23, 24]. Rather than to aim at an effective-site approximation via the $V$-perturbation theory introduced here, a very general result is obtained following the lines of Ref. [20]:

The full conduction-band Green's function $G^c_{\boldsymbol{k}\sigma}(i\omega_l) = 1/[i\omega_l - \varepsilon_{\boldsymbol{k}\sigma} - \Sigma^c_{\boldsymbol{k}\sigma}(i\omega_l)]$ of the Anderson-lattice model at $V_{\boldsymbol{k}} = V$ and (via exact equations of motion) the $T$-matrix

$$T_{\boldsymbol{k}\sigma}(i\omega_l) \equiv |V|^2 F_{\boldsymbol{k}\sigma}(i\omega_l) = \frac{\Sigma^c_{\boldsymbol{k}\sigma}(i\omega_l)}{1 - G^0_{\boldsymbol{k}\sigma}(i\omega_l)\Sigma^c_{\boldsymbol{k}\sigma}(i\omega_l)} \ , \ G^0_{\boldsymbol{k}\sigma}(i\omega_l) = \frac{1}{i\omega_l - \varepsilon_{\boldsymbol{k}\sigma}} \tag{4}$$

are expressed through the band-self energy $\Sigma^c_{\boldsymbol{k}\sigma}$. The latter is expanded in $V$ using bare cumulants [25, 26] which involve only the atomic ($V = 0$ and any $U$) Hamiltonian and bare band-electron Green's functions $G^0_{\boldsymbol{k}\sigma}$. An effective-site approximation is made in the corresponding skeleton expansion [27] (see Fig. 3) through a decomposition of $G^c_{\boldsymbol{k}\sigma} = \mathcal{G}^c_\sigma + \widehat{G}^c_{\boldsymbol{k}\sigma}$, $\mathcal{G}^c_\sigma = \frac{1}{\# sites} \sum_{\boldsymbol{k}} G^c_{\boldsymbol{k}\sigma}$, $\widehat{G}^c_{\boldsymbol{k}\sigma} = G^c_{\boldsymbol{k}\sigma} - \mathcal{G}^c_\sigma$ and the neglect of its non-local part $\widehat{G}^c_{\boldsymbol{k}\sigma}$.

$$\Sigma^c_{\boldsymbol{k}\sigma}(i\omega_l) = \underset{i\omega_l}{\overset{\boldsymbol{k}\;\sigma}{\longleftarrow}}\bullet\underset{i\omega_l}{\overset{\boldsymbol{k}\;\sigma}{\longleftarrow}} + \cdots + \cdots + \cdots + \cdots + \cdots + \mathcal{O}(V^{10})$$

Figure 3: *Skeleton-cumulant expansion of the band-self energy for the Anderson lattice. Full conduction-band lines $G^c_{\boldsymbol{k}\sigma}$ connect bare atomic cumulants (filled circles), which are k- (and σ-) conserving and local, i.e. k-independent. They carry a $V^{(*)}$ at each incoming (outgoing) band line, also if amputated.*

Now $\Sigma^c_{\boldsymbol{k}\sigma}[G^c_{\boldsymbol{k}'\sigma'}] \to \Sigma^{imp}_\sigma[\mathcal{G}^c_{\sigma'}]$ is the local (k-independent) band-self energy of an effective impurity model, and as a consequence the relations

$$\mathcal{G}^c_\sigma = 1/\left[(\widetilde{G}^c_\sigma)^{-1} - \Sigma^{imp}_\sigma\right] \;,\; \mathcal{T}_\sigma = \frac{1}{\# sites}\sum_{\boldsymbol{k}} T_{\boldsymbol{k}\sigma} = \Sigma^{imp}_\sigma/\left[1 - \widetilde{G}^c_\sigma \Sigma^{imp}_\sigma\right] \quad (5)$$

hold by means of an expansion of $\mathcal{G}^c_\sigma$ and the local T-matrix $\mathcal{T}_\sigma$ in real space on the lattice. The propagator $\widetilde{G}^c_{\mu\mu',\sigma} = \delta_{\mu,\mu'}\widetilde{G}^c_\sigma$ emerging here is defined as

$$\widetilde{G}^c_\sigma = G^0_{\mu\mu,\sigma} + \sum_{\nu_1 \neq \mu} G^0_{\mu\nu_1,\sigma}\Sigma^{imp}_{\nu_1\nu_1,\sigma}G^0_{\nu_1\mu,\sigma} + \sum_{\nu_1,\nu_2 \neq \mu} G^0_{\mu\nu_1,\sigma}\Sigma^{imp}_{\nu_1\nu_1,\sigma}G^0_{\nu_1\nu_2,\sigma}\Sigma^{imp}_{\nu_2\nu_2,\sigma}G^0_{\nu_2\mu,\sigma} + \ldots$$

with $G^0_{\mu\mu',\sigma} = \frac{1}{\# sites}\sum_{\boldsymbol{k}} e^{i\boldsymbol{k}(\boldsymbol{R}_\mu - \boldsymbol{R}_{\mu'})} G^0_{\boldsymbol{k}\sigma}$, $\Sigma^{imp}_{\mu\mu',\sigma} = \delta_{\mu,\mu'}\Sigma^{imp}_\sigma$ and is according to (5) to be identified as the effective-impurity's 'bare' band propagator. It can be determined from (4) and (5) resulting in a lattice $T$-matrix and self-consistency constraint as known from the XNCA ([22], Eqn. (2.25), (2.21)):

$$T_{\bm{k}\sigma} = 1/\left[(\mathcal{T}_\sigma)^{-1} - [G^0_{\bm{k}\sigma} - \widetilde{G}^c_\sigma]\right] \;,\; 1 = \frac{1}{\#\,sites}\sum_{\bm{k}} 1/\left[1 - \mathcal{T}_\sigma[G^0_{\bm{k}\sigma} - \widetilde{G}^c_\sigma]\right] \;. \tag{6}$$

Since the expansion that leads to (6) containes no reference to any specific way bare (atomic) cumulants are to be calculated [26], they were only used as a means, the 'outer XNCA-loop' (6) represents a general and well defined effective-site approximation. In this framework the present approach with 'unprojected' Green's functions in Feynman diagrams extends to the Anderson lattice.

In the limit of infinite dimensions ($d \to \infty$) [23] the local skeleton-self energy $\Sigma^{imp}_\sigma$ becomes exact following the arguments of Ref. [24], which proves (6) an exact mapping scheme in this limit. Via equations of motion it can be shown to be equivalent to the 'usual' $d \to \infty$-mapping scheme [23] and the one introduced in [28].

In conclusion, a way to set up an expansion of Green's functions around the atomic limit of the Anderson-impurity model via conventional Feynman diagrams has been demonstrated in considering the requirements for Wick's theorem without 'slave' particles introduced. Important diagram classes have been identified within an analytical treatment of the Kondo problem, indicating the way to self-consistent conserving approximations. In order to obtain an extension to the Anderson-lattice model, however, we had to refer to the effective-site approximation by way of a general cumulant expansion around the atomic limit of the lattice. It remains the question whether correct spectral functions of the spin-degenerate Anderson lattice are thereby obtainable in the low temperature regime [22, 28]. This is under consideration via numerical investigation of conserving approximations mentioned above.

**Acknowledgements:**

The author is very grateful to Prof. N. Grewe for many intensive discussions on the subject of this work and a careful reading of the manuscript.